\documentclass[11pt]{article}
\usepackage{fancyvrb,url,graphicx}
\usepackage{amsmath}
\usepackage{eso-pic}

\newcommand{\code}[1]{{\small\texttt{#1}}}
\newcommand{\dfn}[1]{\emph{#1}}
\newcommand{\pkg}[1]{{#1}}
\DefineVerbatimEnvironment{example}{Verbatim}{fontsize=\footnotesize}

\begin{document}
\AddToShipoutPictureFG*{\AtPageUpperLeft{
   \raisebox{-0.75in}{\makebox[1.5in]{}\parbox[t]{6in}{\small This article has been accepted for publication in \emph{The R Journal}, Volume 8 (2016).}
}}}

\title{\pkg{statmod}: Probability Calculations for the Inverse Gaussian Distribution}
\author{G\"{o}knur Giner and Gordon K.\ Smyth\\
Walter and Eliza Hall Institute of Medical Research}

\date{2 February 2016\thanks{With minor revisions 24 April 2016}}

\maketitle

\begin{abstract}
The inverse Gaussian distribution (IGD) is a well known and often used probability distribution for which fully reliable numerical algorithms have not been available.
Our aim in this article is to develop software for this distribution for the R programming environment.
We develop fast, reliable basic probability functions (\code{dinvgauss}, \code{pinvgauss}, \code{qinvgauss} and \code{rinvgauss}) that work for all possible parameter values and which achieve close to full machine accuracy.
The most challenging task is to compute quantiles for given cumulative probabilities and we develop a simple but elegant mathematical solution to this problem.
We show that Newton's method for finding the quantiles of a IGD always converges monotonically when started from the mode of the distribution.
Simple Taylor series expansions are used to improve accuracy on the log-scale.
The IGD probability functions provide the same options and obey the same conventions as do probability functions provided in the standard R \pkg{stats} package.
The IGD functions are part of the \pkg{statmod} package available from the CRAN repository.
\end{abstract}

\section{Introduction}

The inverse Gaussian distribution (IGD) \cite{tweedie1957inversegaussian,johnson1970continuous} is widely used in a variety of application areas including reliability and survival analysis \cite{whitmore1975inversegauss,chhikara1977invgausslifetime,bardsley1980inversegauss,chhikara1989inversegauss,wang2010inverse,balakrishna2014inverse}.
It is more generally used for modeling non-negative positively skewed data because of its connections to exponential families and generalized linear models \cite{seshadri1993inversegauss,blough1999modeling,smyth1999adjusted,dejong2008glms}.

Our aim in this article is to develop reliable software for this distribution for the R programming environment (\url{http://www.r-project.org}).
Basic probability functions for the IGD have been implemented previously in James Lindsey's R package \pkg{rmutil} \cite{rmutil} and in the CRAN packages \pkg{SuppDists} \cite{SuppDists} and \pkg{STAR} \cite{STAR}.
We have found however that none of these IGD functions work for all parameter values or return results to full machine accuracy.
Bob Wheeler remarks in the \pkg{SuppDists} documentation that the IGD ``is an extremely difficult distribution to treat numerically''.
The \pkg{rmutil} package was removed from CRAN in 1999 but is still available from Lindsey's webpage (\url{http://www.commanster.eu/rcode.html}).
\pkg{SuppDists} was orphaned in 2013 but is still available from CRAN.
The \pkg{SuppDists} code is mostly implemented in C while the other packages are pure R as far as the IGD functions are concerned.

The probability density of the IGD has a simple closed form expression and so is easy to compute.
Care is still required though to handle infinite parameter values that correspond to valid limiting cases.
The cumulative distribution function (cdf) is also available in closed form via an indirect relationship with the normal distribution \cite{shuster1968inverse,chhikara1974estimation}.
Considerable care is nevertheless required to compute probabilities accurately on the log-scale, because the formula involves a sum of two normal probabilities on the un-logged scale.
Random variates from IGDs can be generated using a combination of chisquare and binomial random variables \cite{michael1976generating}.
Most difficult is the inverse cdf or quantile function, which must be computed by some iterative numerical approximation.

Two strategies have been used to compute IGD quantiles.
One is to solve for the quantile using a general-purpose equation solver such as the \code{uniroot} function in R.
This is the approach taken by the \code{qinvgauss} functions in the \pkg{rmutil} and \pkg{STAR} packages.
This approach can usually be relied on to converge satisfactorily but is computationally slow and provides only limited precision.
The other approach is to use Newton's method to solve the equation after applying an initial approximation \cite{kallioras2014percentile}.
This approach was taken by one of the current authors when developing inverse Gaussian code for S-PLUS \cite{smyth1998invgauss}.
It is also the approach taken by the \code{qinvGauss} function in the \pkg{SuppDists} package.
This approach is fast and accurate when it works but can fail unpredictably when the Newton iteration diverges.
Newton's method cannot in general be guaranteed to converge, even when the initial approximation is close to the required value, and the parameter values for which divergence occurs are hard to predict.

We have resolved the above difficulties by developing a Newton iteration for the IGD quantiles that has guaranteed convergence.
Instead of attempting to find a starting value that is close to the required solution, we instead use the convexity properties of the cdf function to approach the required quantiles in a predictable fashion.
We show that Newton's method for finding the quantiles of an IGD always converges when started from the mode of the distribution.
Furthermore the convergence is monotonic, so that backtracking is eliminated.
Newton's method is eventually quadratically convergent, meaning that the number of decimal places corrected determined tends to double with each iteration \cite{press1992numericalrecipes}.
Although the starting value may be far from the required solution, the rapid convergence means the starting value is quickly left behind.
Convergence tends to be rapid even when the required quantile in the extreme tails of the distribution.

The above methods have been implemented in the \code{dinvgauss}, \code{pinvgauss}, \code{qinvgauss} and \code{rinvgauss} functions of the \pkg{statmod} package \cite{statmod}.
The functions give close to machine accuracy for all possible parameter values.
They obey similar conventions to the probability functions provided in the \pkg{stats} package that is bundled with R.
Tests show that the functions are faster, more accurate and more reliable than existing functions for the IGD.
Every effort has to made to ensure that the functions return results for the widest possible range of parameter values.

\section{Density function}
\label{sec:density}

The inverse Gaussian distribution, denoted IG($\mu$,$\phi$), has probability density function (pdf)
\begin{equation}
d(x;\mu,\phi)=\left(2\pi\phi x^3\right)^{-1/2} \exp\left\{-\frac{(x-\mu)^2}{2\phi\mu^2 x}\right\}
\label{pdf}
\end{equation}
for $x>0$, $\mu>0$ and $\phi>0$.
The mean of the distribution is $\mu$ and the variance is $\phi\mu^3$.
In generalized linear model theory \cite{mccullagh1989glms,smyth1999adjusted}, $\phi$ is called the \dfn{dispersion} parameter.
Another popular parametrization of the IGD uses $\lambda=1/\phi$, which we call the \dfn{shape} parameter.
For best accuracy, we compute $d(x;\mu,\phi)$ on the log-scale and then exponentiate if an unlogged value is required.

Note that the mean $\mu$ can be viewed as a scaling parameter: if $X$ is distributed as IG($\mu$,$\phi$), then $X/\mu$ is also inverse Gaussian with mean $1$ and dispersion $\phi\mu$.
The skewness of the distribution is therefore determined by $\phi\mu$, and in fact $\phi\mu$ is the squared coefficient of variation of the distribution. 

\begin{figure}[t]
\begin{center}
\includegraphics[width=\textwidth]{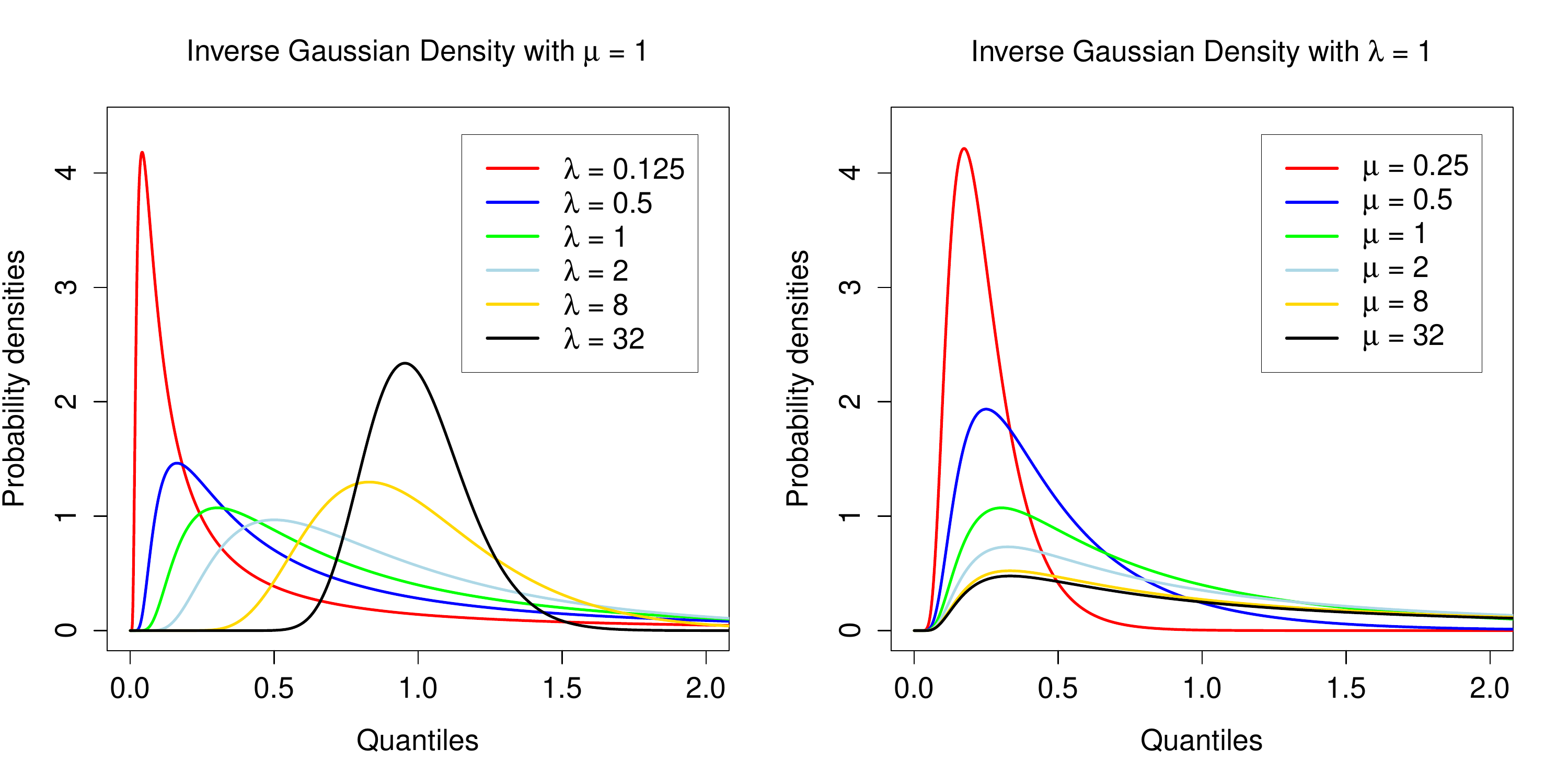}
\caption{Probability density functions of inverse Gaussian distributions.
The left panel shows densities for different $\lambda$ with $\mu=1$. 
The right panel shows densities for different $\mu$ for $\lambda=1$. 
The densities are unimodal with mode between 0 and $\mu$.
As $\mu/\lambda$ increases the distribution becomes more right skew and the mode decreases relative to the mean.
Note that $\lambda=1/\phi$.}
\label{fig:pdf}
\end{center}
\end{figure}

The IGD is unimodal with mode at
\begin{equation}
m=\mu\left\{\left(1+\kappa^2\right)^{1/2}-\kappa\right\}
\label{eq:mode}
\end{equation}
where $\kappa=3\phi\mu/2$ \cite{johnson1970continuous}.
The second factor in the mode is strictly between 0 and 1, showing that the mode is strictly between 0 and $\mu$.
Figure~\ref{fig:pdf} shows the pdf of the IGD for various choices of $\mu$ and $\lambda$.

\begin{table}
\begin{center}
\begin{tabular}{lcccc}
\hline
Description & Parameter values & log-pdf & pdf & cdf\\
\hline
Left limit & $x<0$ & $-\infty$ & 0 & 0\\
Left limit & $x=0$, $\mu>0$ and $\phi<\infty$ & $-\infty$ & 0 & 0\\
Left limit & $x<\mu$ and $\phi=0$ & $-\infty$ & 0 & 0\\
Right limit & $x=\infty$ & $-\infty$ & 0 & 1\\
Right limit & $x>\mu$ and $\phi=0$ & $-\infty$ & 0 & 1\\
Right limit & $x>0$ and $\phi=\infty$ & $-\infty$ & 0 & 1\\
Spike & $x=\mu<\infty$ and $\phi=0$ & $\infty$ & $\infty$ & 1\\
Spike & $x=0$ and $\phi=\infty$ & $\infty$ & $\infty$ & 1\\
Inverse chisquare & $\mu=\infty$ and $\phi<\infty$ & Eqn \ref{pdfinfmean} & Eqn \ref{pdfinfmean} & Uses \code{pchisq}\\
Invalid & $\mu<0$ or $\phi<0$ & \code{NA} & \code{NA} & \code{NA}\\
\hline
\end{tabular}
\caption{Probability density function values for special cases of the parameter values.
The pdf values for infinite parameters are theoretical limit values.}
\label{tab:specialcases}
\end{center}
\end{table}

Care needs to be taken with special cases when evaluating the pdf (Table~\ref{tab:specialcases}).
When $\phi\mu$ is large, a Taylor series expansion shows that the mode becomes dependent on $\phi$ only:
\begin{equation}
m
=\mu\kappa\left\{\left(1+\kappa^{-2}\right)^{1/2}-1\right\}
=\mu\kappa\left(\frac{1}{2\kappa^2}-\frac{1}{8\kappa^4}+\frac{1}{16\kappa^6}-\cdots\right)
\approx \mu\kappa\frac{1}{2\kappa^2}
=\frac{1}{3\phi}.
\label{eq:modetaylor}
\end{equation}
Under the same conditions, the peak value of the density can be seen to converge to $\phi (2\pi/27)^{-1/2}$ $\times\exp(-3/2)$.
This shows that the distribution has a spike at 0 whenever $\phi$ is very large, regardless of $\mu$.
It is also known that
\begin{equation}
\frac{(X-\mu)^2}{\phi X \mu^2} \sim \chi^2_1
\label{chisq}
\end{equation}
\cite{shuster1968inverse}.
Amongst other things, this implies that $1/(X\phi) \sim \chi^2_1$ asymptotically for $\mu$ large.
For infinite $\mu$, the density becomes
\begin{equation}
d(x;\infty,\phi)=\left(2\pi x^3 \phi\right)^{-1/2} \exp\left(-\frac{1}{2\phi x}\right).
\label{pdfinfmean}
\end{equation}
The pdf is always \code{NA} if $x$ is \code{NA}.
Missing values for $\phi$ lead to \code{NA} values for the pdf except when $x<0$ or $x=\infty$.
Missing values for $\mu$ lead to \code{NA} values for the pdf except when $x<0$, $x=\infty$ or $\phi=\infty$.

Next we give some code examples.
We start by loading the packages that we will compare.
Note that \pkg{statmod} is loaded last and is therefore first in the search path.
\begin{example}
> library(rmutil)
> library(SuppDists)
> library(STAR)
> library(statmod)
\end{example}
The \pkg{statmod} \code{dinvgauss} function checks for out-of-range or missing values:
\begin{example}
> options(digits = 3)
> dinvgauss(c(-1, 0, 1, 2, Inf, NA), mean = 1.5, dispersion = 0.7)
[1] 0.000 0.000 0.440 0.162 0.000    NA
\end{example}
Infinite mean corresponds to an inverse-chisquare case:
\begin{example}
> dinvgauss(c(-1, 0, 1, 2, Inf, NA), mean = Inf, dispersion = 0.7)
[1] 0.000 0.000 0.233 0.118 0.000    NA
\end{example}
Infinite dispersion corresponds to a spike at 0 regardless of the mean:
\begin{example}
> dinvgauss(c(-1, 0, 1, 2, Inf, NA), mean = NA, dispersion = Inf)
[1]   0 Inf   0   0   0  NA
\end{example}
Extreme $x$ values have zero density regardless of the mean or dispersion:
\begin{example}
> dinvgauss(c(-1, 0, 1, Inf), mean = NA, dispersion = NA)
[1]  0 NA NA  0
\end{example}
All the existing functions \code{rmutil::dinvgauss}, \code{SuppDist::dinvGauss} and \code{STAR::dinvgauss} return errors for the above calls; they do not tolerate \code{NA} values, or infinite parameter values, or $x$ values outside the support of the distribution.

\section{Cumulative distribution function}

Let $p(q;\mu,\phi)=P(X\le q)$ be the left tail cdf, and write $\bar p(q;\mu,\phi)$ for the right tail probability $P(X> q)=1-p(q;\mu,\phi)$.
The formula developed by \cite{shuster1968inverse} for the cdf is
\[
p(q;\mu,\phi)=p_{\rm norm}((q_m-1)/r)+\exp{(2/\phi_m)} p_{\rm norm}(-(q_m+1)/r)
\]
where $q_m=q/\mu$, $\phi_m=\phi\mu$, $r=(q\phi)^{1/2}$ and $p_{\rm norm}$ is the cdf of the standard normal distribution.
The right tail probability can be written similarly:
\[
\bar p(q;\mu,\phi)=\bar p_{\rm norm}((q_m-1)/r)-\exp{(2/\phi_m)} p_{\rm norm}(-(q_m+1)/r)
\]
where $\bar p_{\rm norm}$ is the right tail of the standard normal.
The fact that this formula is additive on the unlogged scale poses some numerical problems.
The $p_{\rm norm}()$ evaluations are subject to floating underflow, the $\exp()$ evaluation is subject to overflow, and there is the danger of subtractive cancellation when computing the right tail probability.

It is possible to derive an asymptotic expression for the right tail probability.
If $q$ is very large then:
\[
\log\bar p(q;1,\phi)
\approx \frac{1}{\phi_m} - 0.5\log\pi - \log(2\phi_m) - 1.5\log\left(\frac{q_m}{2\phi_m}+1\right) -\frac{q_m}{2\phi_m}.
\]
See the Appendix for the derivation of this approximation.
This approximation is very accurate when $\phi_m^{-1/2}(q_m-1) > 10^5$, but only gives 2--3 significant figures correctly for more modest values such as $\phi_m^{-1/2}(q_m-1) = 10$.

To avoid or minimize the numerical problems described above, we convert the terms in the cdf to the log-scale and remove a common factor before combining the two term terms to get $\log p$.
Given a quantile value $q$, we compute the corresponding $\log p$ as follows:
\begin{align*}
a &= \log p_{\rm norm}((q_m-1)/r)\\
b &= 2/\phi_m + \log p_{\rm norm} (-(q_m+1)/r)\\
\log p &= a+{\rm log1p}(\exp(b-a))
\end{align*}
where $\log p_{\rm norm}()$ is computed by \code{pnorm} with \code{lower.tail=TRUE} and \code{log.p=TRUE}.
Note also that \code{log1p()} is an R function that computes the logarithm of one plus its argument avoiding subtractive cancellation for small arguments.
The computation of the right tail probability is similar but with
\begin{align*}
a &= \log \bar p_{\rm norm}((q_m-1)/r)\\
\log\bar p &= a + {\rm log1p}(-\exp(b-a)).
\end{align*}
Because of this careful computation, \code{statmod::pinvgauss} function is able to compute correct cdf values even in the far tails of the distribution:
\begin{example}
> options(digits = 4)
> pinvgauss(0.001, mean = 1.5, disp = 0.7)
[1] 3.368e-312
> pinvgauss(110, mean = 1.5, disp = 0.7, lower.tail = FALSE)
[1] 2.197e-18
\end{example}
None of the existing functions can distinguish such small left tail probabilities from zero:
\begin{example}
> rmutil::pinvgauss(0.001, m = 1.5, s = 0.7)
[1] 0
> SuppDists::pinvGauss(0.001, nu = 1.5, lambda = 1/0.7)
[1] 0
> STAR::pinvgauss(0.001, mu = 1.5, sigma2 = 0.7)
[1] 0
\end{example}
\code{rmutil::pinvgauss} doesn't compute right tail probabilities.
\code{STAR::pinvgauss} does but can't distinguish right tail probabilities less than \code{1e-17} from zero:
\begin{example}
> STAR::pinvgauss(110, mu = 1.5, sigma2 = 0.7, lower.tail = FALSE)
[1] 0
\end{example}
\code{SuppDists::pinvGauss} returns non-zero right tail probabilities, but these are too large by a factor of 10:
\begin{example}
> SuppDists::pinvGauss(110, nu = 1.5, lambda = 1/0.7, lower.tail = FALSE)
[1] 2.935e-17
\end{example}

The use of log-scale computations means that \code{statmod::pinvgauss} can accurately compute log-probabilities that are too small to be represented on the unlogged scale:
\begin{example}
> pinvgauss(0.0001, mean = 1.5, disp = 0.7, log.p = TRUE)
[1] -7146.914
\end{example}
None of the other packages can compute log-probabilities less than about $-700$.

\code{pinvgauss} handles special cases similarly to \code{dinvgauss} (Table~\ref{tab:specialcases}).
Again, none of the existing functions do this:
\begin{example}
> pinvgauss(c(-1, 0, 1, 2, Inf, NA), mean = 1.5, dispersion = 0.7)
[1] 0.0000 0.0000 0.5009 0.7742 1.0000     NA
\end{example}
Infinite mean corresponds to an inverse-chisquare case:
\begin{example}
> pinvgauss(c(-1, 0, 1, 2, Inf, NA), mean = Inf, dispersion = 0.7)
[1] 0.000 0.000 0.232 0.398 1.000    NA
\end{example}
Infinite dispersion corresponds to a spike at 0 regardless of the mean:
\begin{example}
> pinvgauss(c(-1, 0, 1, 2, Inf, NA), mean = NA, dispersion = Inf)
[1]  0  1  1  1  1 NA
\end{example}
Extreme $x$ values have cdf equal to 0 or 1 regardless of the mean or dispersion:
\begin{example}
> pinvgauss(c(-1, 0, 1, Inf), mean = NA, dispersion = NA)
[1]  0 NA NA  1
\end{example}

We can test the accuracy of the cdf functions by comparing to the cdf of the $\chi^2_1$ distribution.
For any $q_1<\mu$, let $q_2>\mu$ be that value satisfying
$$z=\frac{(q_1-\mu)^2}{\phi\mu^2 q_1}=\frac{(q_2-\mu)^2}{\phi\mu^2 q_2}.$$
From equation~\ref{chisq}, we can conclude that the upper tail probability for the $\chi^2_1$ distribution at $z$ should be the sum of the IGD tail probabilities for $q_1$ and $q_2$, i.e.,
\begin{equation}
\bar p_{\mathrm chisq}(z)=p(q_1;\mu,\phi)+\bar p(q_2;\mu,\phi).
\label{chisqcdf}
\end{equation}
The following code implements this process for an illustrative example with $\mu=1.5$, $\phi=0.7$ and $q_1=0.1$.
First we have to solve for $q_2$:
\begin{example}
> options(digits = 4)
> mu <- 1.5
> phi <- 0.7
> q1 <- 0.1
> z <- (q1 - mu)^2 / (phi * mu^2 * q1)
> polycoef <- c(mu^2, -2 * mu - phi * mu^2 * z, 1)
> q <- Re(polyroot(polycoef))
> q
[1]  0.1 22.5
\end{example}
The chisquare cdf value corresponding to the left hand size of equation~\ref{chisqcdf} is:
\begin{example}
> options(digits = 18)
> pchisq(z, df = 1, lower.tail = FALSE)
[1] 0.00041923696954098788
\end{example}
Now we compute the right hand size of equation~\ref{chisqcdf} using each of the IGD packages, starting with \pkg{statmod}:
\begin{example}
> pinvgauss(q[1], mean = mu, disp = phi) +
+ pinvgauss(q[2], mean = mu, disp = phi, lower.tail = FALSE)
[1] 0.00041923696954098701
> rmutil::pinvgauss(q[1], m = mu, s = phi) +
+ 1 - rmutil::pinvgauss(q[2], m = mu, s = phi)
[1] 0.00041923696954104805
> SuppDists::pinvGauss(q[1], nu = mu, lambda = 1/phi) +
+ SuppDists::pinvGauss(q[2], nu = mu, lambda = 1/phi, lower.tail = FALSE)
[1] 0.00041923696954101699
> STAR::pinvgauss(q[1], mu = mu, sigma2 = phi) +
+ STAR::pinvgauss(q[2], mu = mu, sigma2 = phi, lower.tail = FALSE)
[1] 0.00041923696954100208
\end{example}
It can be seen that the \pkg{statmod} function is the only one to agree with \code{pchisq} to 15 significant figures, corresponding to a relative error of about $10^{-15}$.
The other three packages give 12 significant figures, corresponding to relative errors of slightly over $10^{-12}$.

More extreme tail values give even more striking results.
We repeat the above process now with $q_1=0.01$:
\begin{example}
> q1 <- 0.01
> z <- (q1 - mu)^2 / (phi * mu^2 * q1)
> polycoef <- c(mu^2, -2 * mu - phi * mu^2 * z, 1)
> q <- Re(polyroot(polycoef))
\end{example}
The reference chisquare cdf value is:
\begin{example}
> pchisq(z, df = 1, lower.tail = FALSE)
[1] 1.6427313604456241e-32
\end{example}
This can be compared to the corresponding values from the IGD packages:
\begin{example}
> pinvgauss(q[1], mean = mu, disp = phi) +
+ pinvgauss(q[2], mean = mu, disp = phi, lower.tail = FALSE)
[1] 1.6427313604456183e-32
> rmutil::pinvgauss(q[1], m = mu, s = phi) +
+ 1 - rmutil::pinvgauss(q[2], m = mu, s = phi)
[1] 0
> SuppDists::pinvGauss(q[1], nu = mu, lambda = 1/phi) +
+ SuppDists::pinvGauss(q[2], nu = mu, lambda = 1/phi, lower.tail = FALSE)
[1] 8.2136568022278466e-33
> STAR::pinvgauss(q[1], mu = mu, sigma2 = phi) +
+ STAR::pinvgauss(q[2], mu = mu, sigma2 = phi, lower.tail = FALSE)
[1] 1.6319986233795599e-32
\end{example}
It can be seen from the above that \pkg{rmutil} and \pkg{SuppDists} do not agree with \code{pchisq} to any significant figures, meaning that the relative error is close to 100\%, while \pkg{STAR} manages 3 significant figures.
\pkg{statmod} on the other hand continues to agree with \code{pchisq} to 15 significant figures.

\section{Inverting the cdf}

Now consider the problem of computing the quantile function $q(p;\mu,\phi)$.
The quantile function computes $q$ satisfying $P(X\le q)=p$.

If $q_n$ is an initial approximation to $q$, then Newton's method is a natural choice for refining the estimate.
Newton's method gives the updated estimate as
$$q_{n+1}=q_n+\frac{p-p(q_n;\mu,\phi)}{d(q_n;\mu,\phi)}.$$
For right-tail probabilities, the Newton step is almost the same:
$$q_{n+1}=q_n-\frac{p-\bar p(q_n;\mu,\phi)}{d(q_n;\mu,\phi)}$$
where now $P(X> q)=p$.
Newton's method is very attractive because it is quadratically convergent if started sufficiently close to the required value.
It is hard however to characterize how close the starting value needs to be to achieve convergence and in general there is no guarantee that the Newton iteration will not diverge or give impossible values such as $q<0$ or $q=\infty$.
Our approach is to derive simple conditions on the starting values such that the Newton iteration always converges and does so without any backtracking.
We call this behavior \dfn{monotonic convergence}.

Recall that the IGD is unimodal for all parameter values with mode $m$ given previously.
It follows that the pdf $d(q;\mu\phi)$ is increasing for all $q<m$ and decreasing for all $q>m$ and the cdf $p(q;\mu,\phi)$ is convex for $q<m$ and concave for $q>m$.
In other words, the cdf has a point of inflexion at the mode of the distribution.

\begin{figure}
\begin{center}
\includegraphics[width=\textwidth]{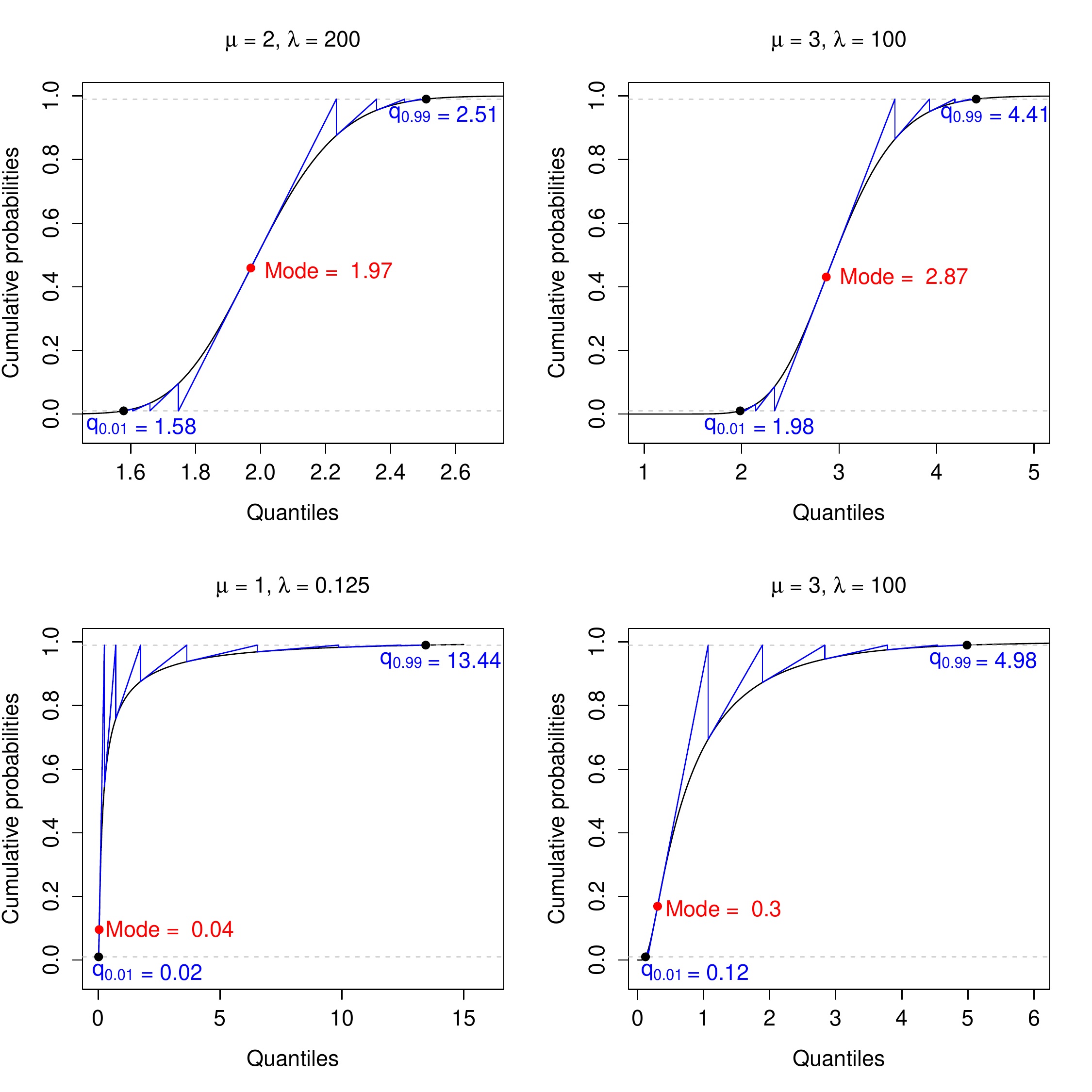}
\caption{Monotonic Newton's method for quantiles of inverse Gaussian distributions.
The cdf has a point of inflexion, marked by a red dot, at the mode of the distribution. 
Blue lines show the progress of the iteration for the 0.01 or 0.99 quantiles. 
Since the cdf is convex to the left of the mode and concave to the right, starting the iteration at the point of inflexion ensures convergence to the required quantiles without any backtracking.}
\label{fig:newton}
\end{center}
\end{figure}

Suppose that the required $q$ satisfies $q \ge m$ and suppose that the working estimate satisfies $m \le q_n \le q$.
It can be seen that the cdf is concave in the interval $[q_n,q]$, the Newton step will be positive and the updated estimate $q_{n+1}$ will still satisfy $m \le q_{n+1} \le q$ (Figure~\ref{fig:newton}).
Suppose instead that $q<m$ and suppose that the working estimate satisfies $q \le q_n \le m$.
In this case it can be seen that the cdf is convex in the interval $[q_n,q]$, the Newton step will be negative and the updated estimate $q_n$ will still satisfy $q \le q_{n+1} \le m$ (Figure~\ref{fig:newton}).
It follows that Newton's method is always monotonically convergent provided that the starting value lies between the mode $m$ and the required value $q$.
In fact the mode $m$ itself can be used as the starting value.
Note that to compute the mode $m$ accurately without subtractive cancellation we use equation~\ref{eq:modetaylor} when $\kappa$ is large and use equation~\ref{eq:mode} otherwise.

We use $q_0=m$ as the starting value for the Newton iteration unless the left or right tail probability is very small.
When the left tail probability is less than $10^{-5}$, we use instead
$$q_0=\frac{\mu}{\phi q_{\rm norm}^2}$$
where $q_{\rm norm}$ is the corresponding quantile of the standard normal distribution.
When the right tail probability is less than $10^{-5}$, we use
$$q_0=q_{\rm gamma}$$
where $q_{\rm gamma}$ is the corresponding quantile of the gamma distribution with the same mean and variances as the IGD.
These starting values are closer to the required $q$ than is $m$ but still lie between $m$ and the required $q$ and so are in the domain of monotonic convergence.
We use the alterative starting values only for extreme tail probabilities because in other cases the computational cost of computing the starting value is greater than the saving enjoyed by reducing the number of Newton iterations that are needed.

The term $p-p(q_n;\mu,\phi)$ in the Newton step could potentially suffer loss of floating point precision by subtractive cancellation when $p$ and $p(q_n;\mu,\phi)$ are nearly equal or if $p$ is very close to 1.
To avoid this we work with $p$ on the log-scale and employ a Taylor series expansion when $p$ and $p(q_n;\mu,\phi)$ are relatively close.
Let $\delta=\log p - \log p(q_n;\mu,\phi)$.
When $|\delta|<10^{-5}$, we approximate
\[
p-p(q_n;\mu,\phi)\approx \delta \exp\left\{\log p + {\rm log1p}(-\delta/2)\right\}.
\]
Here $\log p(q_n;\mu,\phi)$ is computed by \code{pinvgauss} with \code{log.p=TRUE} and ${\rm log1p}(-\delta/2)$ is computed using the \code{log1p} function.

We find that the \pkg{statmod} \code{qinvgauss} package gives 16 significant figures whereas the other packages give no more than 6--8 figures of accuracy.
Precision can be demonstrated by comparing the probability vector $p$ with the values obtained by passing the probabilities through \code{qinvgauss} and \code{pinvgauss}.
\code{qinvgauss} and \code{pinvgauss} are inverse functions, so the final probabilities should be equal in principle to the original values.
Error is measured by comparing the original and processed probability vectors:
\begin{example}
> p <- c(0.000001, 0.00001, 0.0001, 0.001, 0.01, 0.1, 0.5,
+        0.9, 0.99, 0.999, 0.9999, 0.99999, 0.999999)
> 
> p1 <- pinvgauss(qinvgauss(p, mean = 1, disp = 1), mean = 1, disp = 1)
> p2 <- rmutil::pinvgauss(rmutil::qinvgauss(p, m = 1, s = 1), m = 1, s = 1)
> p3 <- SuppDists::pinvGauss(SuppDists::qinvGauss(p, nu = 1, la = 1), nu = 1, la = 1)
> p4 <- STAR::pinvgauss(STAR::qinvgauss(p, mu = 1, sigma2 = 1), mu = 1, sigma2 = 1)
>
> options(digits = 4)
> summary( abs(p-p1) )
    Min.  1st Qu.   Median     Mean  3rd Qu.     Max. 
0.00e+00 0.00e+00 0.00e+00 1.92e-17 2.20e-19 2.22e-16 
> summary( abs(p-p2) )
    Min.  1st Qu.   Median     Mean  3rd Qu.     Max. 
0.00e+00 5.10e-09 8.39e-08 3.28e-07 5.92e-07 1.18e-06 
> summary( abs(p-p3) )
    Min.  1st Qu.   Median     Mean  3rd Qu.     Max. 
1.00e-12 6.00e-12 2.77e-10 1.77e-09 2.58e-09 1.03e-08 
> summary( abs(p-p4) )
    Min.  1st Qu.   Median     Mean  3rd Qu.     Max. 
0.00e+00 0.00e+00 1.20e-08 8.95e-07 2.17e-07 6.65e-06 
\end{example}
It can be seen that the error for \code{statmod::qinvgauss} is never greater than \code{2e-16}.

Similar results are observed if relative error is assessed in terms of the quantile $q$ instead of the probability $p$:
\begin{example}
> q <- qinvgauss(p, mean = 1, disp = 1)
> q1 <- qinvgauss(pinvgauss(q, mean = 1, disp = 1), mean = 1, disp = 1)
> q2 <- rmutil::qinvgauss(rmutil::pinvgauss(q, m = 1, s = 1), m = 1, s = 1)
> q3 <- SuppDists::qinvGauss(SuppDists::pinvGauss(q, nu = 1, la = 1), nu = 1, la = 1)
> q4 <- STAR::qinvgauss(STAR::pinvgauss(q, mu = 1, sigma2 = 1), mu = 1, sigma2 = 1)
> summary( abs(q1-q)/q )
    Min.  1st Qu.   Median     Mean  3rd Qu.     Max. 
0.00e+00 0.00e+00 0.00e+00 5.57e-17 0.00e+00 4.93e-16 
> summary( abs(q2-q)/q )
    Min.  1st Qu.   Median     Mean  3rd Qu.     Max. 
0.00e+00 1.70e-06 3.30e-06 8.94e-05 8.80e-05 5.98e-04 
> summary( abs(q3-q)/q )
    Min.  1st Qu.   Median     Mean  3rd Qu.     Max. 
1.09e-08 3.94e-08 4.78e-08 4.67e-08 5.67e-08 8.93e-08 
> summary( abs(q4-q)/q )
    Min.  1st Qu.   Median     Mean  3rd Qu.     Max. 
0.00e+00 3.00e-07 1.40e-06 9.20e-05 9.42e-05 5.46e-04 
\end{example}
The relative error for \code{statmod::qinvgauss} is never worse than \code{5e-16}.

Speed was determined by generating $p$ as a vector of a million random uniform deviates, and running the \code{qinvgauss} or \code{qinvGauss} functions on p with mean and dispersion both equal to one.
\begin{example}
> set.seed(20140526)
> u <- runif(1000)
> p <- runif(1e6)
> system.time(q1 <- qinvgauss(p, mean = 1, shape = 1))
   user  system elapsed 
   4.29    0.41    4.69 
> system.time(q2 <- rmutil::qinvgauss(p, m = 1, s = 1))
   user  system elapsed 
 157.39    0.03  157.90 
> system.time(q3 <- SuppDists::qinvGauss(p, nu = 1, lambda = 1))
   user  system elapsed 
  13.59    0.00   13.68 
> system.time(q4 <- STAR::qinvgauss(p, mu = 1, sigma2 = 1))
   user  system elapsed 
 266.41    0.06  267.25 
\end{example}
Timings shown here are for a Windows laptop with a 2.7GHz Intel i7 processor running 64-bit R-devel (built 31 January 2016).
The \pkg{statmod} qinvgauss function is 40 times faster than the \pkg{rmutil} or \pkg{STAR} functions about 3 times faster than \pkg{SuppDists}.

Reliability is perhaps even more crucial than precision or speed.
\code{SuppDists::qinvGauss} fails for some parameter values because Newton's method does not converge from the starting values provided:
\begin{example}
> options(digits = 4)
> SuppDists::qinvGauss(0.00013, nu=1, lambda=3)
Error in SuppDists::qinvGauss(0.00013, nu = 1, lambda = 3) : 
Iteration limit exceeded in NewtonRoot()
\end{example}
By contrast, \code{statmod::qinvgauss} runs successfully for all parameter values because divergence of the algorithm is impossible:
\begin{example}
> qinvgauss(0.00013, mean = 1, shape = 3)
[1] 0.1504
\end{example}

\code{qinvgauss} returns right tail values accurately, for example:
\begin{example}
> qinvgauss(1e-20, mean = 1.5, disp = 0.7, lower.tail = FALSE)
[1] 126.3
\end{example}
The same probability can be supplied as a left tail probability on the log-scale, with the same result:
\begin{example}
> qinvgauss(-1e-20, mean = 1.5, disp = 0.7, log.p = TRUE)
[1] 126.3
\end{example}
Note that \code{qinvgauss} returns the correct quantile in this case even though the left tail probability is not distinguishable from 1 in floating point arithmetic on the unlogged scale.
By contrast, the \pkg{rmutil} and \pkg{STAR} functions do not compute right tail values and the \pkg{SuppDists} function fails to converge for small right tail probabilities:
\begin{example}
> SuppDists::qinvGauss(1e-20, nu = 1.5, lambda = 1/0.7, lower.tail = FALSE)
Error in SuppDists::qinvGauss(1e-20, nu = 1.5, lambda = 1/0.7, lower.tail = FALSE) : 
Infinite value in NewtonRoot()
\end{example}
Similarly for log-probabilities, the \pkg{rmutil} and \pkg{STAR} functions do not accept log-probabilities and the \pkg{SuppDists} function gives an error:
\begin{example}
> SuppDists::qinvGauss(-1e-20, nu = 1.5, lambda = 1/0.7, log.p=TRUE)
Error in SuppDists::qinvGauss(-1e-20, nu = 1.5, lambda = 1/0.7, log.p = TRUE) : 
Infinite value in NewtonRoot()
\end{example}

All the \pkg{statmod} IGD functions allow variability to be specified either by way of a dispersion ($\phi$) or shape ($\lambda$) parameter:
\begin{example}
> args(qinvgauss)
function (p, mean = 1, shape = NULL, dispersion = 1, lower.tail = TRUE, 
    log.p = FALSE, maxit = 200L, tol = 1e-14, trace = FALSE) 
\end{example}
Boundary or invalid \code{p} are detected:
\begin{example}
> options(digits = 4)
> qinvgauss(c(0, 0.5, 1, 2, NA))
[1] 0.0000 0.6758    Inf     NA     NA
\end{example}
as are invalid values for $\mu$ or $\phi$:
\begin{example}
> qinvgauss(0.5, mean = c(0, 1, 2))
[1]     NA 0.6758 1.0285
\end{example}

The \pkg{statmod} functions \code{dinvgauss}, \code{pinvgauss} and \code{qinvgauss} all preserve the attributes of the first input argument provided that none of the other arguments have longer length.
For example, \code{qinvgauss} will return a matrix if \code{p} is a matrix:
\begin{example}
> p <- matrix(c(0.1, 0.6, 0.7, 0.9), 2, 2)
> rownames(p) <- c("A", "B")
> colnames(p) <- c("X1", "X2")
> p
      X1     X2
A 0.6001 0.3435
B 0.4919 0.4987
> qinvgauss(p)
      X1     X2
A 0.8486 0.4759
B 0.6637 0.6739
\end{example}
Similarly the names of a vector are preserved on output:
\begin{example}
> p <- c(0.1, 0.6, 0.7, 0.9)
> names(p) <- LETTERS[1:4]
> qinvgauss(p)
     A      B      C      D 
0.2376 0.8483 1.0851 2.1430 
\end{example}

\section{Random deviates}

The functions \code{statmod::rinvgauss}, \code{SuppDists::rinvGauss} and \code{STAR::rinvgauss} all use the same algorithm to compute random deviates from the IGD.
The method is to generate chisquare random deviates corresponding to $(X-\mu)^2/(\phi X \mu^2)$, and then choose between the two possible $X$ values leading to the same chisquare value with probabilities worked out by \cite{michael1976generating}.
The \pkg{SuppDists} function is faster than the others because of the implementation in C.
Nevertheless, the pure R \pkg{statmod} and \pkg{STAR} functions are acceptably fast.
The \pkg{statmod} function generates a million random deviates in about a quarter of a second of elapsed time on a standard business laptop computer while \pkg{STAR} takes about half a second.

The \code{rmutil::rinvgauss} function generates random deviates by running \code{qinvgauss} on random uniform deviates.
This is far slower and less accurate than the other functions.

\section{Discussion}

Basic probability calculations for the IGD have been available in various forms for some time but the functions described here are the first to work for all parameter values and to return close to full machine accuracy.

The \pkg{statmod} functions achieve good accuracy by computing probabilities on the log-scale where possible.
Care is given to handle special limiting cases, including some cases that have not been previously described.
The \pkg{statmod} functions trap invalid parameter values, provide all the standard arguments for probability functions in the R and preserve argument attributes on output.

A new strategy has been described to invert the cdf using a monotonically convergent Newton iteration.
It may seem surprising that we recommend starting the iteration from the same value regardless of the quantile required.
Intuitively, a starting value that is closer to the required quantile might have been expected to be better.
However using an initial approximation runs the risk of divergence, and convergence of Newton's method from the mode is so rapid that the potential advantage of a closer initial approximation is minimized.
The \pkg{statmod} \code{qinvgauss} function is 40 times faster than the quantile functions in the \pkg{rmutil} or \pkg{STAR} packages, despite returning 16 rather than 6 figures of accuracy.
It is also 3 times faster than \pkg{SuppDists}, even though \code{SuppDists::qinvGauss} is written in C, uses the same basic Newton strategy and has a less stringent stopping criterion.
The starting values for Newton's method used by \code{SuppDists::qinvGauss} are actually closer to the final values than those used by \code{statmod::qinvgauss}, but the latter are more carefully chosen to achieve smooth convergence without backtracking.
\code{SuppDists::qinvGauss} uses the log-normal approximation of \cite{whitmore1978normalizing} to start the Newton iteration and the \code{STAR::qinvgauss} uses the same approximation to setup the interval limits for \code{uniroot}.
Unfortunately the log-normal approximation has much heavier tails than the IGD, meaning that the starting values are more extreme than the required quantiles and are therefore outside the domain of monotonic convergence.

As well as the efficiency gained by avoiding backtracking, monotonic convergence has the advantage that any change in sign of the Newton step is a symptom that the limits of floating point accuracy have been reached.
In the \pkg{statmod} \code{qinvgauss} function, the Newton iteration is stopped if this change of sign occurs before the convergence criterion is achieved.

The current \pkg{statmod} functions could be made faster by reimplementing in C, but the pure R versions have benefits in terms of understandability and easy maintenance, and they are only slightly slower than comparable functions such as \code{qchisq} and \code{qt}.

This strategy used here to compute the quantile could be used for any continuous unimodal distribution,
or for continuous distribution that can be transformed to be unimodal.

\begin{example}
> sessionInfo()
R Under development (unstable) (2016-01-31 r70055)
Platform: x86_64-w64-mingw32/x64 (64-bit)
Running under: Windows 7 x64 (build 7601) Service Pack 1

locale:
[1] LC_COLLATE=English_Australia.1252  LC_CTYPE=English_Australia.1252   
[3] LC_MONETARY=English_Australia.1252 LC_NUMERIC=C                      
[5] LC_TIME=English_Australia.1252    

attached base packages:
[1] stats     graphics  grDevices utils     datasets  methods   base     

other attached packages:
 [1] statmod_1.4.24    STAR_0.3-7        codetools_0.2-14  gss_2.1-5        
 [5] R2HTML_2.3.1      mgcv_1.8-11       nlme_3.1-124      survival_2.38-3  
 [9] SuppDists_1.1-9.2 rmutil_1.0       

loaded via a namespace (and not attached):
[1] Matrix_1.2-3    splines_3.3.0   grid_3.3.0      lattice_0.20-33
\end{example}


\newpage
\section*{Appendix: asymptotic right tail probabilities}

Here we derive an asymptotic expression for the right tail probability, $\bar p(q;\mu,\phi)$, when $q$ is large.
Without loss of generality, we will assume $\mu=1$.
First, we drop the $1/x$ term in the exponent of the pdf (\ref{pdf}), leading to:
\[
d(x;1,\phi)\approx \left(2\pi\phi x^3\right)^{-1/2} \exp\left(-\frac{x}{2\phi}+\frac{1}{\phi}\right)
\]
for $x$ large.
Integrating the pdf gives the right tail probability as:
\[
\bar p(q;1,\phi)
\approx \exp\left(\phi^{-1}\right) (2\pi\phi)^{-1/2} \int_q^\infty x^{-3/2} \exp\left(-\frac{x}{2\phi}\right) dx
\]
for $q$ large.
Transforming the variable of integration gives:
\[
\bar p(q;1,\phi)
\approx \exp\left(\phi^{-1}\right) (2\pi\phi)^{-1/2} (2\phi)^{-1/2} \int_{q/(2\phi)}^\infty x^{-3/2} \exp(-x) dx.
\]
Finally, we approximate the integral using
\[
\int_a^\infty x^{-3/2} \exp(-x)dx \approx (a+1)^{-3/2} \exp(-a),
\]
which gives
\[
\bar p(q;1,\phi)
\approx \exp\left(\phi^{-1}\right) \pi^{-1/2} \left(2\phi\right)^{-1} \left(\frac{q}{2\phi}+1\right)^{-3/2} \exp\left(-\frac{q}{2\phi}\right)
\]
and
\[
\log\bar p(q;1,\phi)
\approx \frac{1}{\phi} - 0.5\log\pi - \log(2\phi) - 1.5\log\left(\frac{q}{2\phi}+1\right) -\frac{q}{2\phi}
\]
for $q$ large.

\end{document}